\documentclass[10pt,conference]{IEEEtran}

\usepackage{research5}

\begin{document}

\title{Superimposed Coded and Uncoded Transmissions of a Gaussian
  Source over the Gaussian Channel}

\author{\authorblockN{Shraga Bross}
\authorblockA{Department of Electrical Engineering\\
Technion Israel Institute of Technology\\
\texttt{shraga@ee.technion.ac.il}}
\and
\authorblockN{Amos Lapidoth ~ ~ ~ Stephan Tinguely}
\authorblockA{Signal and Information Processing Laboratory\\
Swiss Federal Institute of Technology (ETH) Zurich, Switzerland\\
\texttt{ \{lapidoth, tinguely\}@isi.ee.ethz.ch}}}

\maketitle

\begin{abstract}
  We propose to send a Gaussian source over an average-power limited
  additive white Gaussian noise channel by transmitting a linear
  combination of the source sequence and the result of its
  quantization using a high dimensional Gaussian vector quantizer. We
  show that, irrespective of the rate of the vector quantizer (assumed
  to be fixed and smaller than the channel's capacity), this
  transmission scheme is asymptotically optimal (as the quantizer's
  dimension tends to infinity) under the mean squared-error fidelity
  criterion.  This generalizes the classical result of Goblick about
  the optimality of scaled uncoded transmission, which corresponds to
  choosing the rate of the vector quantizer as zero, and the classical
  source-channel separation approach, which corresponds to choosing
  the rate of the vector quantizer arbitrarily close to the capacity
  of the channel.
\end{abstract}

\section{Introduction}
The minimal distortion with which a memoryless source can be
communicated over a memoryless noisy channel is given by the
evaluation at channel capacity of the distortion vs.\ rate function
corresponding to the source law and the fidelity criterion \cite[Thm.
9.6.3]{Gallager68}. For a memoryless Gaussian source and an
average-power limited additive white Gaussian noise (AWGN) channel two
classical schemes are known to achieve this minimum distortion: the
source-channel separation approach \cite{shannon48} and Goblick's
``uncoded'' scheme \cite{goblick65}. (See also \cite{elias67},
\cite{mittal_phamdo2000}, and \cite{Gastpar03}.) Here we shall show
that these two schemes can be viewed as the endpoints of a continuum
of optimal transmission schemes.  In the proposed transmission schemes
the transmitted waveform is a linear combination of the source
sequence and of the result of its quantization using a Gaussian vector
quantizer. The source-channel separation approach corresponds to
having the rate of the vector quantizer be arbitrarily close to
channel capacity, and Goblick's uncoded scheme corresponds to having
the rate of the vector quantizer be zero.

We point out that in contrast to other work on hybrid digital-analog
joint source-channel coding, e.g.~\cite{shamai_verdu_zamir98},
\cite{skoglund02} and \cite{mittal02}, we do not aim for issues like
``robust'' communication, but merely mean to point out a
generalization of two well-known optimal schemes. Also, it should be
emphasized that our transmission schemes do not increase
bandwidth. This should be contrasted with the problem addressed by
Shamai, Verd\'u and Zamir \cite{shamai_verdu_zamir98} where a
memoryless source is to be transmitted to a receiver via \emph{two
  independent channels} where the transmission over one of the
independent channels is required to be uncoded.

\section{Some Definitions}

To state our contribution more precisely we need some definitions. The
additive white Gaussian noise channel is a channel whose time-$k$ output
$Y_{k}$ takes value in the set of reals $\Reals$ and is given by
\begin{equation}
  \label{eq:Gaussin_channel}
  Y_{k} = x_{k} + Z_{k}
\end{equation}
where $x_{k} \in \Reals$ denotes the time-$k$ channel input and where
the random variables $\{Z_{k}\}$ are IID, zero-mean, variance-$N$,
Gaussian random variables. We say that the length-$n$ sequence of
inputs $x_{1}, \ldots, x_{n}$ satisfies the average power constraint
if
\begin{equation}
  \label{eq:avg_power}
  \frac{1}{n} \sum_{i=1}^{n} x_{i}^{2} \leq P.
\end{equation}
The capacity of the additive white Gaussian noise channel under the above
average power constraint is given by
\begin{equation}
  \label{eq:capacity}
  C = \frac{1}{2} \log \left(1 + \frac{P}{N} \right).
\end{equation}
(We assume throughout that $N$ is strictly larger than zero.)

The memoryless zero-mean variance-$\sigma^{2}$ source is a source that
emits the sequence $\{S_{i}\}$ of IID zero-mean
variance-$\sigma^{2}$ Gaussian random variables. The variance
$\sigma^{2}$ is assumed to be strictly larger than zero. The
distortion vs.\ rate function $D(R)$ corresponding to this source and
to the single-letter squared-error fidelity measure $d(s, \hat{s}) =
(s-\hat{s})^{2}$ is given by
\begin{equation}
  \label{eq:DofR}
  D(R) = \sigma^{2} 2^{-2R}.
\end{equation}

The evaluation of the above distortion vs.\ rate function at the
capacity of the additive Gaussian noise channel is given by
\begin{align}
  \label{eq:DofRatC}
  D^{*}  & = \left. \sigma^{2} 2^{-2R} \right|_{R = \frac{1}{2} \log (1 +
    P/N)} \nonumber \\
  & = \sigma^{2} \frac{N}{P + N}. 
\end{align}

A blocklength-$n$ transmission scheme is a pair of mappings
$f_{n},\phi_{n}: \Reals^{n} \rightarrow \Reals^{n}$ with the
understanding that when the source emits the sequence $\vect{s} \in
\Reals^{n}$ the sequence 
\begin{equation*}
  f_{n}(\vect{s}) \triangleq 
  \bigl( x_{1}(\vect{s}), \ldots, x_{n}(\vect{s})\bigr)
\end{equation*}
is fed to the channel. We require that the transmitted sequence
satisfy the average power constraint 
\begin{equation*}
  \E{\|f_{n}(\vect{S})\|^{2}} \leq n \cdot P
\end{equation*}
i.e., 
\begin{equation}
  \label{eq:avg5}
  \frac{1}{n} \sum_{i=1}^{n} \E{x_{i}^{2}(\vect{S})} \leq P. 
\end{equation}
The channel then produces the output sequence $\vect{Y}$ whose $i$-th
component $Y_{i}$ is given by $Y_{i} = x_{i}(\vect{S}) + Z_{i}, \quad
i=1,\ldots, n$. This output sequence is then mapped by $\phi_{n}$ to
the reconstruction sequence $\hat{\vect{S}}$:
\begin{align*}
  \hat{\vect{S}} & = \phi_{n}(\vect{Y}) \\
  & \triangleq \bigl( \hat{S}_{1}(\vect{Y}), \ldots,
  \hat{S}_{n}(\vect{Y}) \bigr).
\end{align*}
The distortion associated with $(f_{n},\phi_{n})$ is given by 
\begin{equation}
  \label{eq:dist_f}
  d(f_{n}, \phi_{n}) = \frac{1}{n} \sum_{i=1}^{n} \E{\bigl( S_{i} -
  \hat{S}_{i}(\vect{Y}) \bigr)^{2}}
\end{equation}
where $S_{i}$ and $\hat{S}_{i}$ denote the $i$-th component of
$\vect{S}$ and $\hat{\vect{S}}$ respectively.

A sequence of schemes $\{f_{n}, \phi_{n}\}$ indexed by the blocklength
$n$ is said to be asymptotically optimal for the transmission of a
Gaussian source over the additive white Gaussian noise channel under the
mean squared-error fidelity criterion if it results in the
transmitted sequence satisfying the average power constraint
\eqref{eq:avg5} i.e., 
\begin{equation}
  \label{eq:avg10}
  \E{\| f_{n}(\vect{S}) \|^{2}} \leq n P
\end{equation}
and if 
\begin{equation}
  \label{eq:def_opt_dist}
  \varlimsup_{n \rightarrow \infty} d(f_{n}, \phi_{n}) = D^{*}.
\end{equation}

In this submission we propose a sequence of asymptotically optimal
transmission schemes parameterized by the free parameter
\begin{equation}
  0 < \rho < \frac{1}{2} \log \left(1 + \frac{P}{N} \right)
\end{equation}
which corresponds to the rate of the Gaussian vector quantizer that we
employ. Thus, to each fixed $\rho$ as above, we present a sequence $\{f_{n},
\phi_{n}\}$ of coding schemes (parameterized by the blocklength-$n$)
that is asymptotically optimal.

\section{The Proposed Scheme}

The proposed scheme is conceptually simple, but this simplicity is
masked by some of the epsilons and deltas involved. For the sake of
clarity and brevity we shall therefore omit these epsilons and deltas
here.

At the heart of the scheme is a rate-$\rho$ Gaussian vector quantizer.
We denote the quantizer's codebook by ${\mathcal C}$ and assume that
its $2^{n \rho}$ codewords are chosen independent of each other, each
being drawn uniformly over a centered sphere in $\Reals^{n}$. The
normalized squared-radius of the sphere is roughly $\sigma^{2} -
\sigma^{2} 2 ^{-2 \rho}$ so that the normalized squared-norm of each
of the codewords in ${\mathcal C}$ is given roughly by
\begin{equation}
  \label{eq:approxRadius}
  \frac{1}{n} \|\vect{u}\|^{2} \approx \sigma^{2} - \sigma^{2} 2^{-2\rho}
\end{equation}
where $n$ denotes the blocklength, $\|\vect{u}\|^{2}$ denotes the sum
of the squares of the components of $\vect{u}$, and where $\sigma^{2}$
is the source's variance. 

Notice that this would be a rate-$\rho$ optimal vector quantizer for
this source and that it would yield a quantization error $\Delta$,
where
\begin{equation}
  \Delta \approx \sigma^{2} 2^{-2\rho}.
\end{equation}
Also, if we slightly increase $\rho$ or slightly decrease the radius of
the sphere on which the codewords of the quantizer lie, we could (with
very high probability) find a codeword $\vect{u}^{*} \in {\mathcal C}$
such that $\vect{s} - \vect{u}^{*}$ would be nearly orthogonal to
$\vect{u}^{*}$.  Such a codeword $\vect{u}^{*}$ would then satisfy
\begin{align}
  n^{-1} \|\vect{s} - \vect{u}^{*}\|^{2} & \approx n^{-1} \|\vect{s}\|^{2}
  - n^{-1} \|\vect{u}^{*}\|^{2} \nonumber \\
  & \approx \sigma^{2} - \bigl( \sigma^{2} - \sigma^{2} 2^{-2 \rho}
  \bigr) \nonumber \\
  \label{eq:approxD}
  & = \sigma^{2} 2^{-2 \rho}
\end{align}
where the first approximation follows because $\vect{s} -
\vect{u}^{*}$ is nearly orthogonal to $\vect{u}^{*}$ and where the
second approximation follows by the law of large numbers for
$\|\vect{S}\|^{2}/n$ and from our choice of the radius of the
quantizer's sphere.

We can now describe the encoding schemes. Observing the source
sequence $\vect{s}$, we choose the codeword $\vect{u}^{*}$ in the vector
quantizer's codebook ${\mathcal C}$
equiprobably among all codewords that have a ``typical'' angle to
$\vect{s}$, i.e.~equiprobably among all $\vect{u} \in \mathcal{C}$
satisfying
\begin{equation}\label{eq:enc_cond}
\inner{\frac{\vect{s}}{\norm{\vect{s}}}}{\frac{\vect{u}}{\norm{\vect{u}}}}
\approx \sqrt{1-2^{-2\rho}},
\end{equation} 
where $\inner{\cdot}{\cdot}$ is the standard inner product in
$\Reals^{n}$. If no such $\vect{u} \in \mathcal{C}$ exists, the
codeword $\vect{u}^{*}$ is choosen to be the all zero sequence
$\vect{0}$. By slightly shrinking the quantizer's sphere we can
guarantee that with very high probability there exists at least one
$\vect{u} \in \mathcal{C}$ satisfying \eqref{eq:enc_cond}, and
consequently, with help of the weak law of large numbers for
$\norm{\vect{S}}/n$, that with very high probability
\begin{equation}
  \label{eq:approx_orth}
  \inner{\vect{s}   - \vect{u}^{*}}{\vect{u}^{*}} \approx 0.
\end{equation}
The transmitted sequence $\vect{x} \triangleq f_{n}(\vect{s})$ is now given
by a linear combination of $\vect{u}^{*}$ and the source sequence
$\vect{s}$:
\begin{equation}
  \vect{x} = f_{n}(\vect{s}) = \alpha \vect{s} + \beta \vect{u}^{*}
\end{equation}
where the coefficients $\alpha = \alpha(\rho)$ and $\beta =
\beta(\rho)$ are judiciously chosen as 
\begin{align}
\beta(\rho) &= \sqrt{\frac{P+N}{\sigma^2}} - \alpha(\rho),\\[2mm]
\alpha(\rho) &= \sqrt{\frac{2^{-2\rho}(N+P)-N}{\sigma^2 2^{-2\rho}}}.
\end{align}
This choice of $\alpha$ and $\beta$ is dictated by two requirements.
The first is that $\vect{x}$ roughly satisfy the power constraint.
Indeed, writing
\begin{equation}
  \vect{x} = (\alpha + \beta) \vect{u}^{*} + \alpha(\vect{s} - \vect{u}^{*})
\end{equation}
we note that by \eqref{eq:approx_orth} we shall have $\frac{1}{n}
\|\vect{x}\|^{2} \approx P$ if
\begin{equation*}
  (\alpha + \beta)^{2} \|\vect{u}\|^{2}/n + \alpha^{2} \|\vect{s} -
  \vect{u}^{*}\|^{2}/n \approx P
\end{equation*}
or, in view of \eqref{eq:approxRadius} and \eqref{eq:approxD}, if
\begin{equation}
  (\alpha + \beta)^{2} \sigma^{2} \bigl( 1 - 2^{-2\rho} \bigr) +
  \alpha^{2} \sigma^{2} 2^{-2\rho} \approx P.
\end{equation}

The second requirement dictating the choice of $\alpha$ and $\beta$
has to do with the decoding and will be described as soon as we
describe how the source sequence is reconstructed from the channel
output.

This reconstruction takes place in two phases. In the first phase the
decoder makes a guess $\hat{\vect{u}} \in {\mathcal C}$ of the
transmitted codeword $\vect{u}^{*} \in {\mathcal C}$. In the second
phase the decoder then makes an estimate $\hat{\vect{s}}$ of the
source sequence based on $\vect{y}$ and $\hat{\vect{u}}$. The guess of
$\vect{u}^{*}$ in the first phase is based on the observation
\begin{equation}
  \label{eq:output}
  \vect{Y} = (\alpha + \beta) \vect{u}^{*} + \alpha (\vect{s} -
  \vect{u}^{*}) + \vect{Z}.
\end{equation}
The decoder treats the scaled quantization noise $\alpha (\vect{s} -
\vect{u}^{*})$ as Gaussian noise, and thus ``sees'' a signal
($(\alpha + \beta) \vect{u}^{*}$) of average power $(\alpha + \beta)
\sigma^{2}(1 - 2^{-2\rho})$ contaminated in additive noise ($\alpha
(\vect{s} - \vect{u}^{*}) + \vect{Z}$) of variance $\alpha^{2}
\sigma^{2} 2^{-2\rho} + N$. Using minimum angle decoding, i.e.~$\hat{\vect{u}}
= \argmax_{\vect{u} \in {\mathcal C}} \inner{\vect{y}}{\vect{u}}$, it can be
shown after some analysis that the decoder will succeed with high
probability if \cite{Lapidoth96}
\begin{equation}
  \rho < \frac{1}{2} \log \left( 
    1 + \frac{(\alpha + \beta) \sigma^{2}(1 - 2^{-2\rho})}
             {\alpha^{2} \sigma^{2}2^{-2\rho} + N} \right).
\end{equation}
Replacing this inequality with an (approximate) equality gives us the
second condition on $\alpha, \beta$.

In the second phase the reconstructor assumes that the first phase was
 successful in identifying the codeword $\vect{u}^{*}$. Rearranging
 terms in \eqref{eq:output} we have
 \begin{equation*}
 \frac{\vect{Y}-(\alpha + \beta) \vect{u}^{*}}{\alpha} = (\vect{s} -
 \vect{u}^{*}) + \frac{1}{\alpha} \vect{Z}.
 \end{equation*}
And, since $\vect{u}^{*}$ and $\vect{s}-\vect{u}^{*}$ are nearly
orthogonal, a reasonable estimator of $\vect{S}$ is now the linear
estimator
\begin{equation}
  \label{eq:Shat}
  \hat{\vect{S}} = \vect{u}^{*} + \frac{\alpha^2 \Delta}{\alpha^2 \Delta
  + N} \cdot \frac{\vect{Y} - (\alpha + \beta)\vect{u}^{*} }{\alpha},
\end{equation}
and this is, indeed, the reconstructor we propose.  Thus, the
reconstruction function $\phi_{n}$ can be formally defined as
\begin{equation}
  \phi_n(\vect{y}) = \hat{\vect{u}} + \frac{\alpha^2 \Delta}{\alpha^2 \Delta
  + N} \cdot \frac{\vect{y} - (\alpha + \beta)\hat{\vect{u}} }{\alpha},
\end{equation}
where $\hat{\vect{u}} = \argmax_{\vect{u} \in {\mathcal C}}
\inner{\vect{y}}{\vect{u}}$ and $\Delta = \sigma^2 2^{-2\rho}$.

The expected squared error associated to the proposed sequence of
schemes $\{f_n,\phi_n\}$
\begin{equation}\label{eq:dist_f_2}
d(f_n,\phi_n) = \frac{1}{n} \E{\norm{\vect{S} - \hat{\vect{S}}}^2},
\end{equation}
where the expectation is taken over all $\vect{S}$, $\vect{Z}$ and
$\mathcal{C}$, can now be analyzed by using
\begin{displaymath}
\vect{s} - \hat{\vect{s}} \approx \frac{1}{\alpha^2 \Delta+N} \left(
  N \vect{s} - \alpha \beta
  \Delta \vect{u}^{\ast} - \alpha \Delta\vect{z} - (N-\alpha\beta
  \Delta)\hat{\vect{u}} \right),
\end{displaymath}
in \eqref{eq:dist_f_2}. Writing the expectation as a sum of the
individual cross-terms (most of which are straightforwardly bounded)
and showing that
\begin{align*}
\E{\inner{\vect{S}}{\vect{U}^{\ast}}} &\gtrapprox n \sigma^2 (1-2^{-2R}),\\ 
\E{\inner{\vect{S}}{\widehat{\vect{U}}}} &\gtrapprox n \sigma^2 (1-2^{-2R}),\\
\E{\inner{\vect{Z}}{\widehat{\vect{U}}}} &\approx 0,
\end{align*}
then results in
\begin{displaymath}
\frac{1}{n} \E{\norm{\vect{S} - \hat{\vect{S}}}^2} \approx
\sigma^2 \frac{N}{N+P}.
\end{displaymath}

Of course, the rigorous analysis also requires analyzing the effect
that the non-existence of a codeword $\vect{u} \in \mathcal{C}$
satisfying the encoder condition \eqref{eq:enc_cond} and the effect
that an error in identifying $\vect{u}^{*}$ entail, as well as
justifying the approximations that we have presented.

\section{Conclusion}
We have shown that for the transmission of an IID Gaussian source
over an AWGN channel with average input power constraint, the minimal
expected squared error distortion can be achieved by the superposition
of coded and uncoded transmission, for arbitrary power repartition
among the schemes. 
The preserved correlation between the source sequence and the
transmitted codeword makes the coded and uncoded schemes perfectly
compatible.

\end{document}